\documentclass[preprint]{revtex4}
\usepackage{graphicx}
\usepackage{subfigure}
\usepackage{multirow}
\usepackage{hyperref}
\usepackage{amsmath,epstopdf,bm,amssymb,amsfonts}
\usepackage{color}
\usepackage{amssymb,amsmath}

\begin{document}
\title{Accelerated expansion of the Universe in the Presence of Dark Matter Pressure}

\author{Zeinab Rezaei\footnote{E-mail: zrezaei@shirazu.ac.ir}}

\affiliation{Department of Physics, Shiraz
University, Shiraz 71454, Iran.\\
Biruni Observatory, Shiraz
University, Shiraz 71454, Iran.}

\begin{abstract}
Expansion dynamics of the Universe is one of the important subjects in modern cosmology.
The dark energy equation of state determines this dynamics so that the Universe is in an accelerating phase.
However, the dark matter can also affect the accelerated expansion of the Universe through its equation of state.
In the present work, we explore the expansion dynamics of the Universe in the presence of dark matter pressure.
In this regard, applying the dark matter equation of state from the observational data related to the
rotational curves of galaxies, we calculate the evolution of dark matter density. Moreover, the Hubble parameter,
history of scale factor, luminosity distance, and deceleration parameter are studied while the dark matter pressure
is taken into account. Our results verify that the dark matter pressure leads to the higher values of the
Hubble parameter at each redshift and the expansion of the Universe grows due to the DM pressure.
\end{abstract}
\maketitle
\section{Introduction}

Type Ia supernova observational data confirm an accelerating phase for the
Universe \cite{Moresco}.
Accelerated expansion of the Universe can be explained considering an
energy component named dark energy (DE) in Friedmann equation derived from general relativity.
Using the multiwavelength observation of blazars, it is possible to measure the expansion rate of
the Universe \cite{Domínguez}.
The capability of different configurations of the space interferometer eLISA to probe the late-time background expansion of the Universe using gravitational
wave standard sirens has been studied \cite{Tamanini}.
Different cosmological models have been constrained to understand the expansion dynamics of the Universe from galaxy cluster scales \cite{Wang}.

Different aspects of the accelerated expansion of the Universe have been investigated
\cite{Russ,Tsamis,Parker,Carroll,Linder,Dehghani,Kolb,Berger,Bamba,Avelino,Balakin,Komatsu,Akarsu,Aguila,Akarsu15,Cesare,Tu}.
The influence of inhomogeneities on the global expansion factor by averaging the Friedmann equation
has been calculated \cite{Russ}.
The effect of various particles such as massless fermions, gauge bosons, and conformally coupled scalars
on the cosmic expansion rate relative to that of the graviton has been explored \cite{Tsamis}.
The vacuum energy of a free quantized field of very low mass may
alter the recent expansion of the Universe \cite{Parker}.
An empirical evidence relating to the Friedmann
equation and the dynamical relation in general relativity between the expansion rate of
the Universe and the energy density have been presented \cite{Carroll}.
The precision of distance-redshift observations indicating the acceleration-deceleration transition and the components and equations of state of the energy density have been studied \cite{Linder}.
The acceleration of the expanding Universe can be explained by Gauss-Bonnet gravity with negative Gauss-Bonnet coefficient and
without a cosmological constant \cite{Dehghani}.
The second order in perturbation
variables the expansion rate of an inhomogeneous Universe and the corrections to the
evolution of the expansion rate have been demonstrated \cite{Kolb}.
A model of interacting DE where the DE density is related by the
holographic principle to the Hubble parameter is in agreement with the observational
data supporting an accelerating Universe \cite{Berger}.
Nonminimal Yang-Mills theory in which the field couples to a function of the scalar curvature
can realize both inflation and the late-time accelerated expansion of the Universe \cite{Bamba}.
Considering a pressureless fluid with a constant bulk viscosity driving the present accelerated
expansion of the Universe, a bulk viscous matter-dominated Universe model has been presented \cite{Avelino}.
Accelerated expansion of the Universe can be driven by traditional matter
with positive pressure because of the back-reaction of the gravity
field \cite{Balakin}.
Considering a non-adiabatic-like accelerated expansion of the Universe in entropic cosmology
shows that the increase of the entropy for the simple model is uniform \cite{Komatsu}.
Distinct behaviors of the scalar and vector fields together with
the real valued mass gained by the Stueckelberg mechanism lead the Universe to go through the
two different accelerated expansion phases with a decelerated expansion phase between them \cite{Akarsu}.
General conditions for the acceleration in the expansion of the Universe and a
solution for the Weyl scalar field describing a cosmological model for the present time have been presented \cite{Aguila}.
Higher dimensional steady state cosmologies
with constant volume for which the three dimensional external space
is expanding at an accelerated rate but the internal space
is contracting have been explored \cite{Akarsu15}.
Effective Friedmann equation from the dynamics of group field theory shows the occurrence of an era of
accelerated expansion without the need to introduce an inflaton field \cite{Cesare}.
The evolution solutions of the FRW Universe have been derived by combination of the Friedmann acceleration equation
based on the thermodynamics of the Hubble horizon and the evolution equation
of the Universe based on the energy balance relation \cite{Tu}.

Dark matter (DM) in the Universe alters the accelerated expansion of the Universe \cite{Avelino10,Balakin11,Kleidis}.
A bulk viscous matter-dominated Universe with a pressureless fluid characterizes
both the baryon and DM components has been considered to
explain the present accelerated expansion of the Universe \cite{Avelino10}.
Applying the Archimedean-type coupling of the DM
with DE to study the late-time accelerated
expansion confirms that the Archimedean-type coupling provides the Universe evolution to be a quasiperiodic and/or multistage process \cite{Balakin11}.
Studying the dynamical
characteristics of a spatially-flat cosmological model in which instead of DE
the DM possesses some sort of fluid-like properties indicates that the pressure becomes negative enough, so that the Universe
accelerates its expansion \cite{Kleidis}.

However, the DM as one of the significant portion of the Universe can affect the astrophysical systems through its pressure
\cite{Bharadwaj,Muller,Faber,Binder,Nakajima,Su,Saxton,Bezares,Wechakama,Guzman,Harko1,Harko2,Bettoni,Barranco,Kunz}.
The observations of rotation curves together with the gravitational lensing can determine the equation of state (EOS) of DM \cite{Bharadwaj}.
The constraints on the EOS of DM have been presented employing CMB, supernovae Ia, and large scale structure data
in a modified $\Lambda$CDM cosmology \cite{Muller}.
The observations of galaxy rotation curves and gravitational lensing give the density and pressure profiles of the galactic fluid \cite{Faber}.
A decelerated-accelerated transition considering the present value of the deceleration parameter can occur
considering the non-vanishing DM pressure and
negative small values of the coupling constant \cite{Binder}.
Degeneracy pressure of fermionic DM affects the flat-top column density profile of clusters of galaxies \cite{Nakajima}.
The non-ideal fluid EOS for the DM halo can be obtained from the observations of gravitational lensing deflection angle
\cite{Su}.
Extended theories of DM considering self-interaction, non-extensive thermostatistics, and
 boson condensation
can be explained using the polytropic EOS of DM haloes \cite{Saxton}.
Pressures related to an EOS parameter of total energy of the same value as for weak fields in solar-relativistic ranges
can be the result of the DM dominance in the scalar-field excitations of induced gravity with a Higgs potential
\cite{Bezares}.
Inner slope of halo density profile and the mass and the annihilation cross-section of DM particles into
electron-positron pairs are influenced by the pressure from DM annihilation \cite{Wechakama}.
The contribution of accreted DM to the supermassive black hole growth is affected by the DM with non trivial pressure near a supermassive black hole \cite{Guzman}.
The radial and tangential pressures of anisotropic DM have been studied considering a mixture of two different non-interacting perfect fluids \cite{Harko1}.
The energy density and the
radial pressure of the DM halos have a general r-dependent functional relationship \cite{Harko2}.
The effective pressure of the DM component, originated from the non-minimal coupling between gravity
and DM, reduces the growth of structures at galactic scales \cite{Bettoni}.
The DM EOS obtained using the observational data from the rotation curves of galaxies has a functional dependence universal
for all galaxies \cite{Barranco}.
The DM pressure has also been constrained by the large-scale cosmological observations \cite{Kunz}.
Since the accelerated expansion of the Universe is influenced by the properties of DM, it is necessary to
study the effects of DM pressure on the expansion dynamics of the Universe. Here, we investigate
the accelerated expansion of the Universe in the presence of the DM pressure.

%-----------------------------------------------------------
\section{Dark Matter density evolution in the Presence of Dark Matter Pressure}

Starting with a homogeneous and isotropic cosmology, the Friedmann equations are,
\begin{eqnarray}\label{f1}
\dot{a}^2+k c^2 =\frac{8\pi G}{3}\rho a^2,
 \end{eqnarray}
\begin{eqnarray}\label{f2}
\dot{a}^2+k c^2+2a \ddot{a} =-\frac{8\pi G}{c^2}P a^2.
 \end{eqnarray}
In the above equations, $a(t)$ is the scale factor and also the total density, $\rho$, is related to the DM
density, $\rho_{DM}$, and DE density, $\rho_{DE}$, by $\rho=\rho_{DM}+\rho_{DE}$. In addition, the total pressure, $P$,
is expressed in terms of the DM pressure, $P_{DM}$, and DE pressure, $P_{DE}$, as $P=P_{DM}+P_{DE}$. Besides, $k=-1,0$, and $1$ for an open, flat, and closed Universe, respectively. With the DM EOS $P_{DM}=0$ and the DE EOS $P_{DE}=-c^2 \rho_{DE}$ in $\Lambda$CDM model, we know that the combination of Eqs. (\ref{f1}) and (\ref{f2}) leads to the following density evolution for DM,
\begin{eqnarray}\label{e1}
\rho_{DM}(a)=\rho_{DM0} a^{-3}.
 \end{eqnarray}
In Eq. (\ref{e1}), $\rho_{DM0}$ denotes the DM density at the present time.
%%%%%%%%%%%%%%%%%%%%%%%%%%%%%%%%%%%%%%%%%%%%
%%%%%%%%%%%%%%%%%%%%%%%%%%%%%%%%%%%%%%%%%%%%%%%%%%%%%%%%%%%%%%%%%%%%%%%%%%%%%%%%%%%%%%%%%%%%%%%%%%%%
However, in this work, we are interested in the effects of the DM pressure on the DM density evolution as well as the accelerated expansion of the Universe. Therefore, we first calculate the density evolution of DM assuming the DM pressure is not zero, i.e. $P_{DM}\neq0$. To do this, we multiply Eq. (\ref{f1})
by $a$ and differentiate both its sides with respect to t. This gives,
\begin{eqnarray}\label{fp1}
\dot{a}(\dot{a}^2+2a\ddot{a}+k c^2) =\frac{8\pi G}{3}\frac{d}{dt}(\rho a^3).
 \end{eqnarray}
Using Eq. (\ref{f2}), the above equation leads to,
\begin{eqnarray}\label{fp2}
\frac{da}{dt}(-\frac{8\pi G}{c^2}P a^2) =\frac{8\pi G}{3}\frac{d}{dt}(\rho a^3),
 \end{eqnarray}
which this also results in,
\begin{eqnarray}\label{fp3}
\frac{d(\rho a^3)}{da} =-\frac{3}{c^2}P a^2.
 \end{eqnarray}
We consider the Eq. (\ref{fp3}) for the DM with the EOS, $P_{DM}=P_{DM}(\rho_{DM})$,
\begin{eqnarray}\label{fp4}
\frac{d(\rho_{DM} a^3)}{da} =-\frac{3}{c^2}P_{DM} a^2.
 \end{eqnarray}
In this paper, we employ the DM EOS obtained from the rotational curves of galaxies \cite{Barranco}.
Applying some simple calculations, Eq. (\ref{fp4}) gives,
\begin{eqnarray}\label{fp5}
\frac{d\rho_{DM}}{\rho_{DM}+P_{DM}/c^2} =-3\frac{da}{a},
 \end{eqnarray}
which can be integrated as follows,
\begin{eqnarray}\label{fp6}
\int_{\rho_{DM0}}^{\rho_{DM}}\frac{d\rho_{DM}}{\rho_{DM}+P_{DM}/c^2} =-3 \int_{a_0}^{a}\frac{da}{a},
 \end{eqnarray}
in which $a_0=1$ is the scale factor at the present time. The left side of Eq. (\ref{fp6}) can be calculated after inserting the DM EOS, i.e. $P_{DM}=P_{DM}(\rho_{DM})$. We present the result of this integration by $B(\rho_{DM})$ to emphasis its dependency on the DM density. Therefore, Eq. (\ref{fp6}) gives,
\begin{eqnarray}\label{fp7}
B(\rho_{DM}) = ln( a^{-3}).
 \end{eqnarray}
Solving the above equation for the value of $\rho_{DM}$ leads to the $a$ dependency of the function $\rho_{DM}(a)$.
In fact, $\rho_{DM}(a)$ describes the DM density evolution.
Eq. (\ref{fp7}) is solved by the fixed point method. In our calculations, the maximum error in the value of $B_{DM}$ is $10^{-7}$.
It should be noted that for the case $P_{DM}=0$, the function $\rho_{DM}(a)$ leads to Eq. (\ref{e1}).

\begin{figure*}
\vspace*{1cm}       % Give the correct figure height in cm
\includegraphics[width=0.36\textwidth]{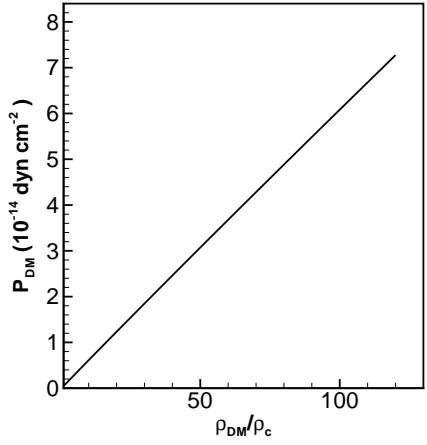}
\caption{Dark matter EOS related to the
galaxy U5750 with the parameters ${\rho}_0=0.31\ GeV/cm^3$ and
$p_0=1.1\times 10^{-8} \ GeV/cm^3$ and $\chi^2_{min}/d.o.f.=0.01$, \cite{Barranco}. $\rho_c$ denotes the critical density of the Universe.}
\label{fig1eos}
\end{figure*}

In this work, we apply the DM EOS obtained using observational data of the rotation curves of galaxies \cite{Barranco}.
The pseudo-isothermal model results in a mass density profile with the property of
regularity at the origin. Using the velocity profile, geometric potentials, and gravitational potential,
the DM EOS obtained applying the pseudo-isothermal density profile has the following form,
\begin{eqnarray}\label{213}
       {P_{DM}}({\rho_{DM}})=\frac{8  {p}_0}{\pi^2-8}[\frac{\pi^2}{8}-\frac{arctan\sqrt{\frac
       {{\rho}_0}{{\rho_{DM}}}-1}}{\sqrt{\frac{{\rho}_0}
       {{\rho_{DM}}}-1}}
	 -\frac{1}{2}(arctan\sqrt{\frac
       {{\rho}_0}{{\rho_{DM}}}-1}\ )^2],
 \end{eqnarray}
in which $\rho_{DM}$ and $P_{DM}$ denote the density and pressure of DM and
the free parameters, $\rho_0$ and $p_0$, are the central density and pressure of galaxies.
This EOS has a functional dependence universal for all galaxies with free parameters which are related with the evolution
history of the galaxy. This universality makes it possible to describe the DM EOS in other scales, e.g. dark-matter admixed
neutron stars \cite{Rezaei} with Eq. (\ref{213}). Here, we suppose that the universality allows the DM EOS to be used in large scale structure
in studying the cosmological accelerated expansion.
The DM EOS which we have used is related to the
galaxy U5750 which is the result of one of the best fit with $\chi^2_{min}/d.o.f.=0.01$ \cite{Barranco}. In Fig. \ref{fig1eos}, we have presented this
DM EOS.

\begin{figure*}
\vspace*{1cm}       % Give the correct figure height in cm
\includegraphics[width=0.7\textwidth]{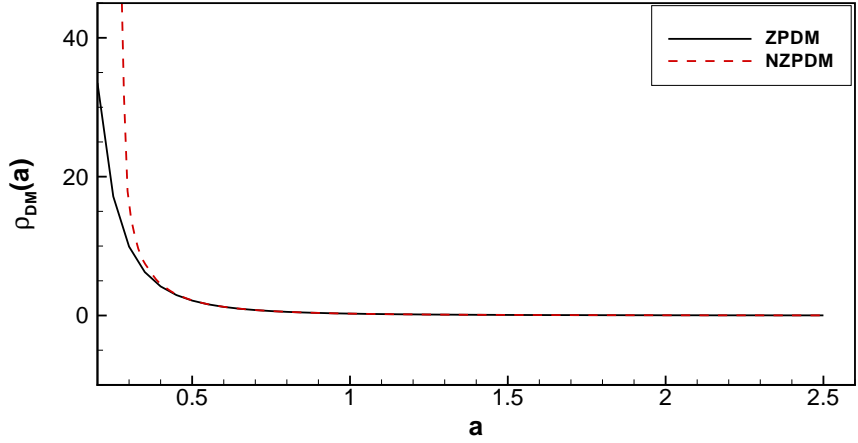}
\caption{Evolution of DM density, $\rho_{DM}(a)$, versus the cosmological scale factor, $a$, for two cases of zero pressure DM (ZPDM) and non zero pressure DM (NZPDM).}
\label{fig1}
\end{figure*}

Fig. \ref{fig1} gives the DM density evolution for both cases of zero pressure DM (ZPDM) and non zero pressure DM (NZPDM).  The evolution of DM density in NZPDM case differs from the ZPDM one.
For lower values of the scale factor, i.e. $a<1$, the existence of the
DM pressure leads to larger DM density. This enhancement is more significant
at $a<0.5$. Besides, at lower values of the cosmological scale factor,
the DM density in the case of NZPDM reduces with scale factor more rapidly than the case of ZPDM. This is while for $a>1$, the density of NZPDM decreases with scale factor
similarly to ZPDM one.

\section{Dynamical expansion of the Universe with Dark Matter Pressure}

Considering a spatially flat universe in which $k=0$ and the Hubble parameter, $H(a)\equiv\dot{a}/a$, Eq. (\ref{f1}) is written as follows,
\begin{eqnarray}\label{fp8}
H^2(a) = \frac{8\pi G}{3}(\rho_{DM}(a)+\rho_{DE}),
 \end{eqnarray}
in which $\rho_{DM}(a)$ is given by Eqs. (\ref{e1}) and (\ref{fp7}) for the cases of $P_{DM}=0$ and $P_{DM}\neq0$, respectively.
\begin{figure*}
\vspace*{1cm}       % Give the correct figure height in cm
\includegraphics[width=0.7\textwidth]{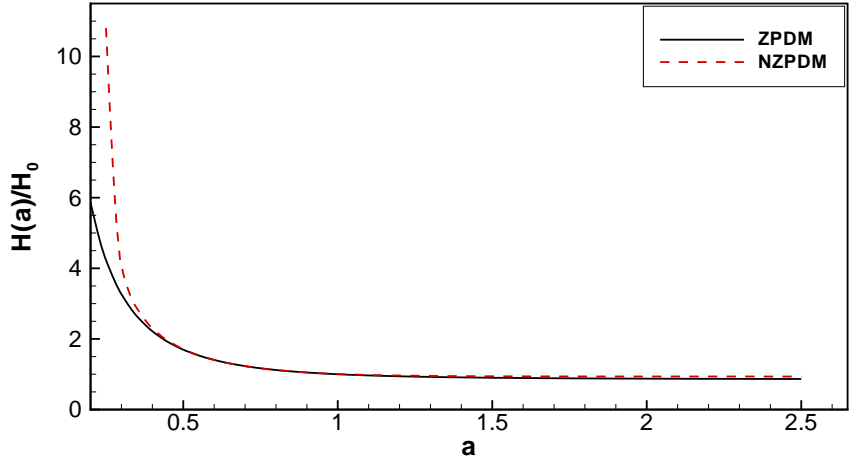}
\caption{Hubble parameter versus the cosmological scale factor, $a$, for two cases of ZPDM and NZPDM. Besides, $H_0$ denotes the present day value of the Hubble parameter.}
\label{fig2}
\end{figure*}
This equation describes the dynamical expansion of the Universe. Fig. \ref{fig2} shows the Hubble parameter as a function of scale factor in two cases of ZPDM and NZPDM. NZPDM predicts higher values for the Hubble parameter, especially at smaller scale factors.
In addition, the rate at which the Hubble parameter decreases with $a$ is affected by the DM pressure.
For $a<1$, the Hubble parameter reduction with scale factor is more considerable when the pressure of DM
is taken into account.

\begin{figure*}
\vspace*{1cm}       % Give the correct figure height in cm
\includegraphics[width=0.65\textwidth]{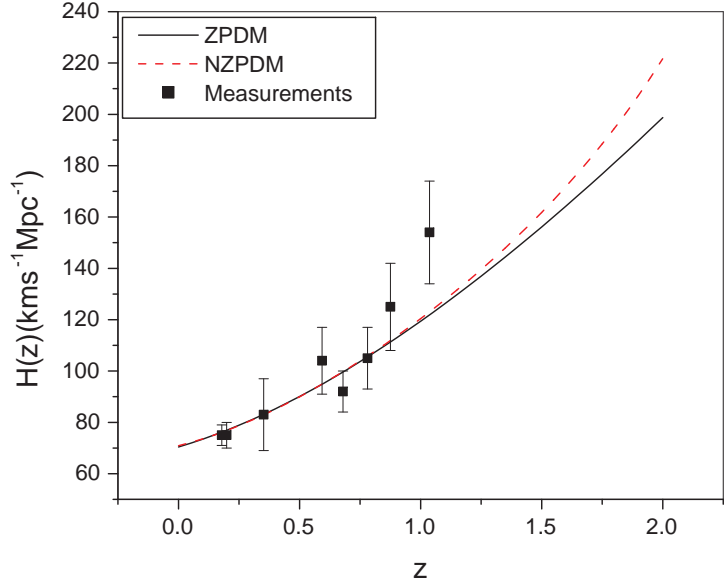}
\caption{Hubble parameter versus the redshift, z, in two cases of ZPDM and NZPDM and the observational data from the median
$D4000_n - z$ relations \cite{Moresco}.}
\label{fig3}
\end{figure*}

\begin{figure*}
\vspace*{1cm}       % Give the correct figure height in cm
\includegraphics[width=0.65\textwidth]{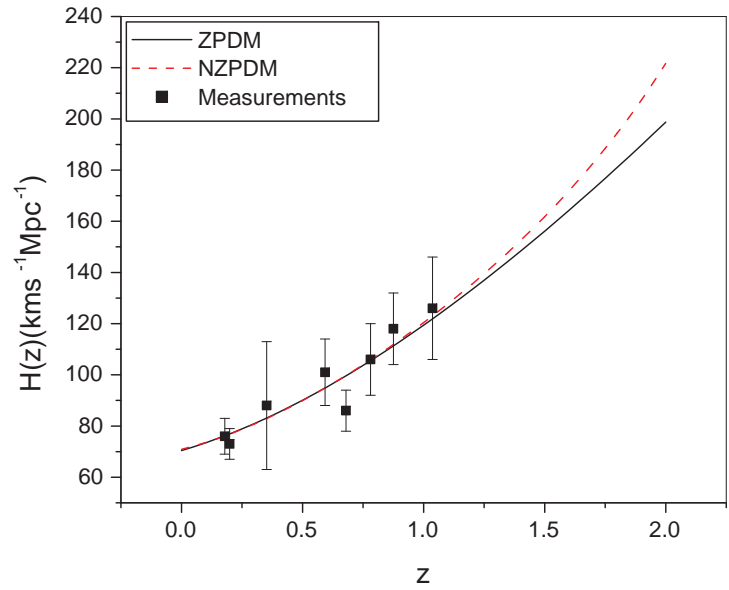}
\caption{Same as Fig. \ref{fig3} but for the observational data from the upper envelope \cite{Moresco}.}
\label{fig4}
\end{figure*}

\begin{figure*}
\vspace*{1cm}       % Give the correct figure height in cm
\includegraphics[width=0.65\textwidth]{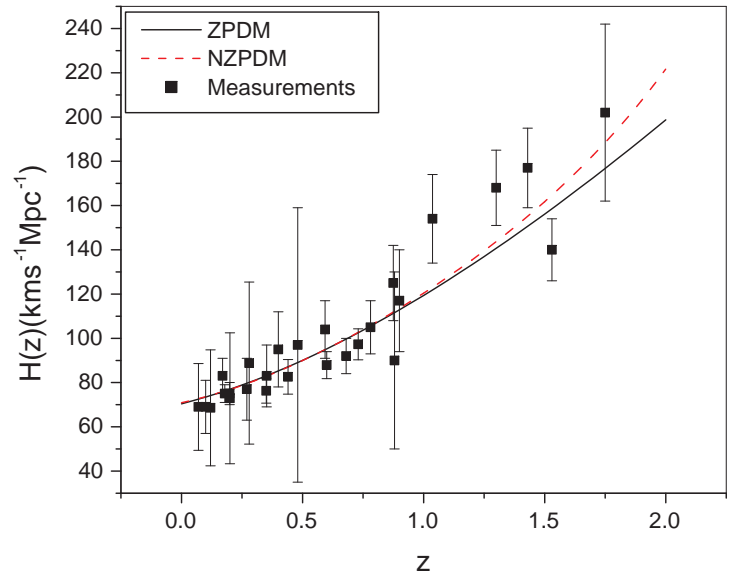}
\caption{Same as Fig. \ref{fig3} but for the observational data from Refs. \cite{Wang} and \cite{Chen}.}
\label{fig20}
\end{figure*}

Figs. \ref{fig3}-\ref{fig20} present the Hubble parameter versus the redshift, z. The redshift is related to the scale factor by $1+z=a^{-1}$  \cite{Weinberg}. Our results have been also compared with the observational data from the median
$D4000_n - z$ relations in Fig. \ref{fig3}, from the upper envelope in Fig. \ref{fig4} \cite{Moresco}, and the observational data from Refs. \cite{Wang} and \cite{Chen} in Fig. \ref{fig20}. At each redshift the Hubble parameter is higher for the case of NZPDM.
Besides, the DM pressure leads to the increase in the growth rate of
the Hubble parameter with the redshift. The Hubble parameter is affected by the
DM pressure more significantly at higher values of the redshift.
Our results for the accelerating
expansion of the Universe with the DM pressure also agree with
the observational data from the median
 $D4000_n - z$ relations and the upper envelope
\cite{Moresco} and the observational data from Refs. \cite{Wang} and \cite{Chen}. Interestingly, for the most observational data, the Hubble parameter is higher than
the theoretical result with ZPDM. This difference can be explained by the DM pressure
which increases the Hubble parameter at each redshift.
In the following, we study the properties of the Universe with the DM pressure.

\subsection{Scale factor a}

\begin{figure*}
\vspace*{1cm}       % Give the correct figure height in cm
\includegraphics[width=0.55\textwidth]{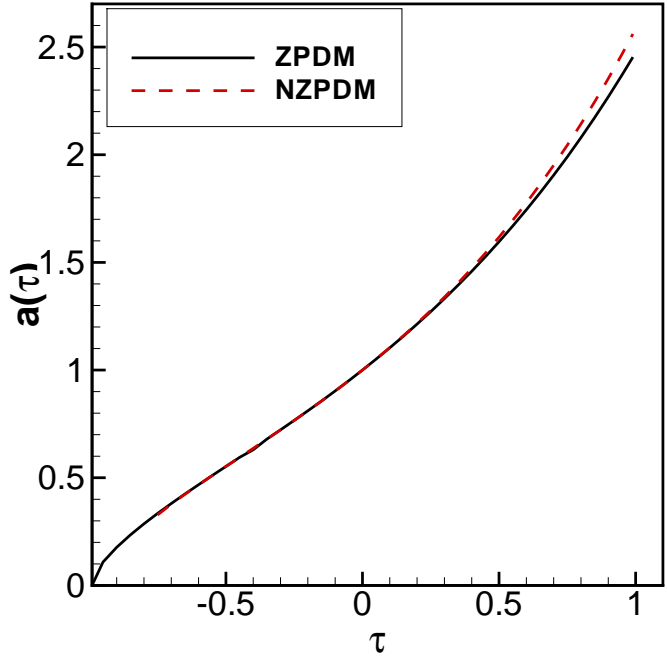}
\caption{Scale factor, $a(\tau)$, versus $\tau=H_0(t-t_0)$ for two cases of ZPDM and NZPDM.}
\label{fig5}
\end{figure*}

\begin{figure*}
\vspace*{1cm}       % Give the correct figure height in cm
\includegraphics[width=0.55\textwidth]{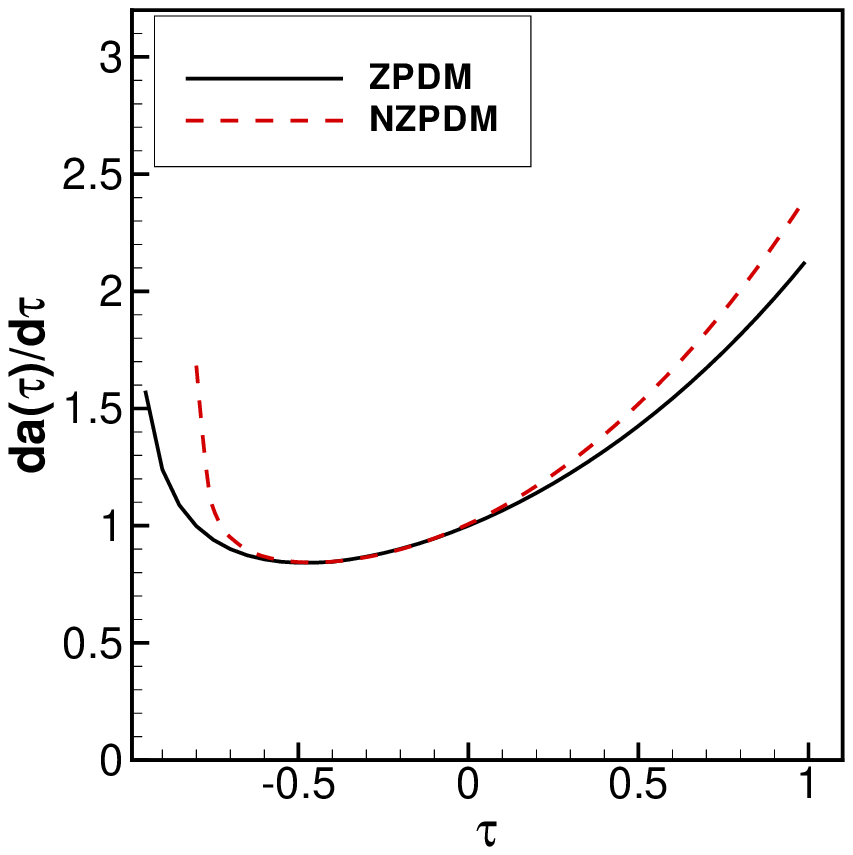}
\caption{Same as Fig. \ref{fig5} but for $d a(\tau)/d\tau$.}
\label{fig6}
\end{figure*}

\begin{figure*}
\vspace*{1cm}       % Give the correct figure height in cm
\includegraphics[width=0.55\textwidth]{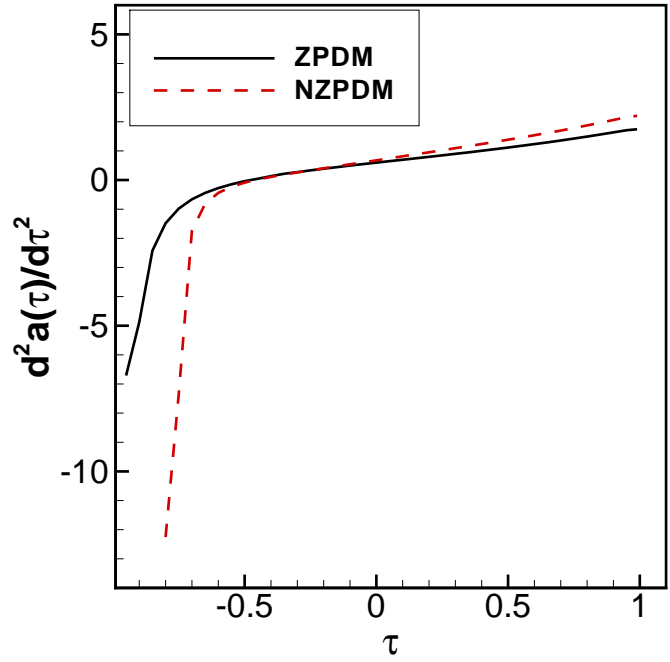}
\caption{Same as Fig. \ref{fig5} but for $d^2 a(\tau)/d\tau^2$.}
\label{fig7}
\end{figure*}

The history of scale factor, $a(t)$, is given by
\begin{eqnarray}\label{at}
\int_1^a\frac{da}{a H(a)}=\int_{t_0}^t dt,
 \end{eqnarray}
in which $t$ denotes the cosmic time and also $t_0$ shows the cosmic time today. Using the definition $\tau=H_0(t-t_0)$, Eq. (\ref{at}) has the following form
\begin{eqnarray}
\int_1^a\frac{H_0da}{a H(a)}=\tau,
 \end{eqnarray}
where $H(a)$ has been given in Eq. (\ref{fp8}). The scale factor, $a(\tau)$, the first derivative of scale factor, $d a(\tau)/d\tau$, and the second derivative of scale factor, $d^2 a(\tau)/d\tau^2$, are presented in Figs. \ref{fig5}-\ref{fig7}.
The increase of the scale factor with the cosmic time is faster if the DM pressure is considered.
This effect is more considerable for $\tau>0$, i.e. after the present time.
Therefore, the expansion of the Universe grows due to the DM pressure. Considering each cosmic time, the value of $d a(\tau)/d\tau$
is higher when the DM pressure is taken into account. The effects of
NZPDM on the $d a(\tau)/d\tau$ are more important when $|\tau|$ is larger.
Moreover, the DM pressure results in the increase of the slope of $d a(\tau)/d\tau$.
It can be seen from Fig. \ref{fig7} that depending on the sign of $d^2 a(\tau)/d\tau^2$,
the effects of NZPDM on this quantity are different. Considering the cosmic times at which
$d^2 a(\tau)/d\tau^2<0$, the second derivative of scale factor decreases with the DM pressure
leading to more negative values for this quantity. However, for the cosmic times with
$d^2 a(\tau)/d\tau^2>0$, the second derivative of scale factor is higher for the case of NZPDM.
Fig. \ref{fig7} also confirms that the DM pressure affects the $d^2 a(\tau)/d\tau^2$ more significantly
when $\tau<0$, i.e. the past time, compared to $\tau>0$.

\subsection{Luminosity distance $d_L$}

\begin{figure*}
\vspace*{1cm}       % Give the correct figure height in cm
\includegraphics[width=0.55\textwidth]{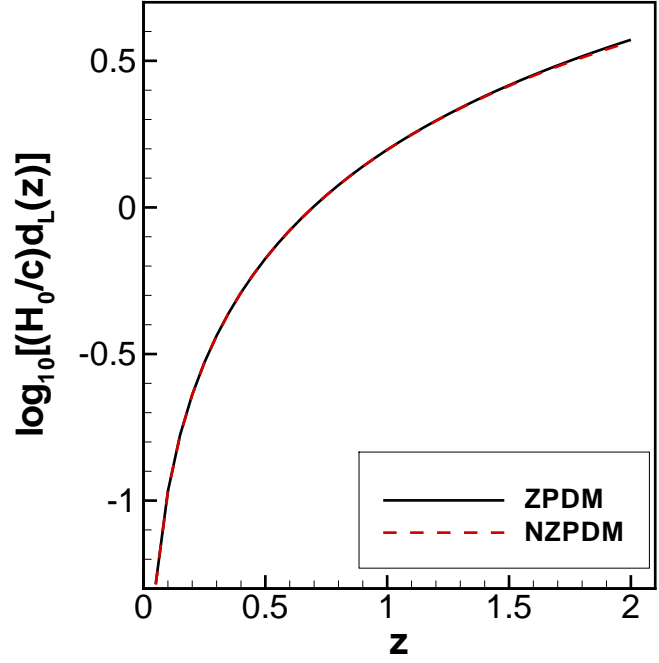}
\caption{Luminosity distance versus the redshift, z, in two cases of ZPDM and NZPDM.}
\label{fig8}
\end{figure*}

\begin{figure}[h!]
	 {\includegraphics[scale=1.05]{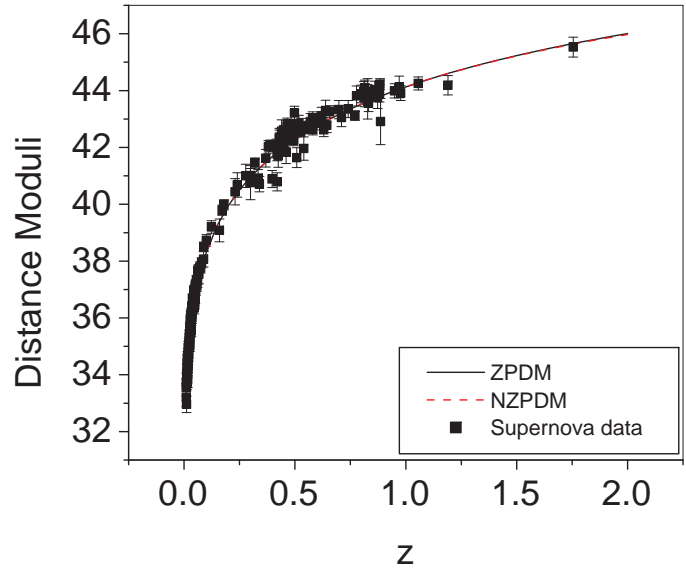}}
	 	\caption{Distance moduli versus the redshift, z, in two cases of ZPDM and NZPDM and the supernova data \cite{Riess4}. }
	\label{fig9}
\end{figure}

One of the important parameters in studying the expansion of the Universe which can be compared with the observational results is the luminosity distance, $d_L$. This quantity is calculated as follows \cite{Riess},
\begin{eqnarray}
d_L(z)=c(1+z)\int_0^{z}\frac{dz}{H(z)}.
 \end{eqnarray}
In addition, considering the same absolute magnitude $M$ for the supernovae, the extinction-corrected distance moduli is as follows \cite{Berger},
\begin{eqnarray}
\mu(z)=5 log_{10}(d_L(z)/Mpc) + 25.
 \end{eqnarray}
Figs. \ref{fig8} and \ref{fig9} show the z dependency of the
luminosity distance and extinction-corrected distance moduli, respectively. The supernova data \cite{Riess4} for the distance moduli are also presented in Fig. \ref{fig9}.
The results for the luminosity distance and distance moduli are not significantly affected by
the DM pressure.
Fig. \ref{fig9} confirms that our results for the distance moduli in the case of NZPDM agree with
the supernova data points.

\subsection{Deceleration parameter q}

\begin{figure*}
\vspace*{1cm}       % Give the correct figure height in cm
\includegraphics[width=0.55\textwidth]{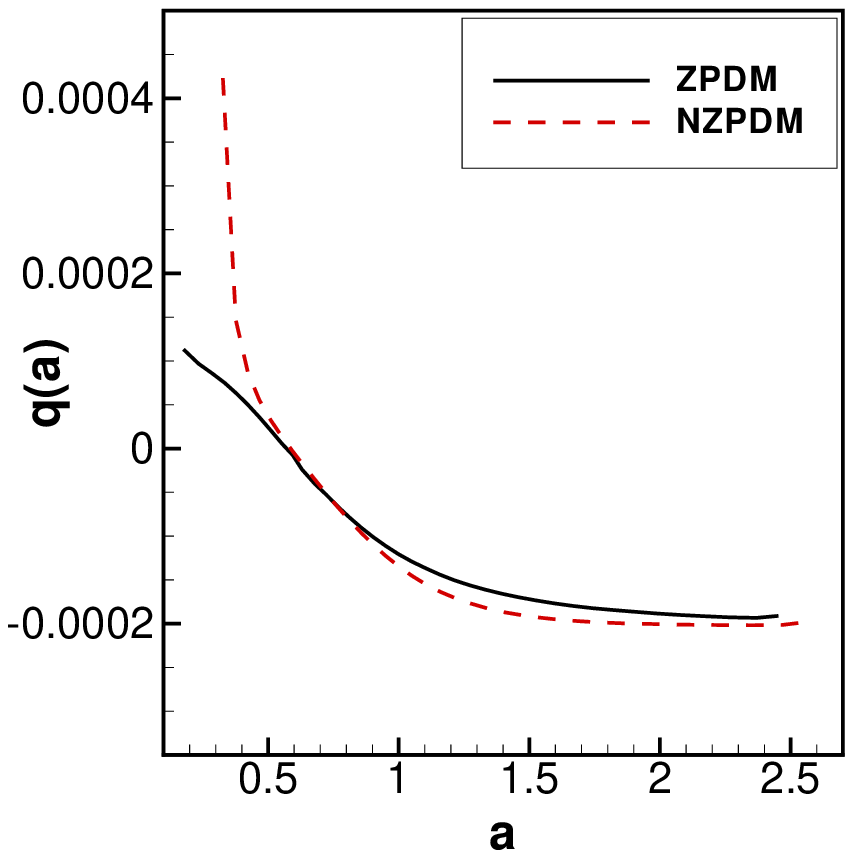}
\caption{Deceleration parameter, $q$, versus the scale factor, $a$, in two cases of ZPDM and NZPDM.}
\label{fig10}
\end{figure*}

The deceleration parameter, $q$, is defined by
\begin{eqnarray}
q(a)=-\frac{\ddot{a}}{a}\frac{1}{H^2}.
 \end{eqnarray}
Fig. \ref{fig10} presents the deceleration parameter as a function of the scale factor, $a$, in two cases of ZPDM and NZPDM.
At lower scale factors, the deceleration is higher if the NZPDM is considered.
However, for $a>1$, the DM pressure leads to more negative values for the deceleration
parameter which this corresponds to more acceleration. The deceleration
parameter is more affected by the DM pressure for $a<1$.

\section{Summary and Conclusions}

Dark matter (DM) equation of state from the observational data of the
rotational curves of galaxies has been employed
to investigate the accelerated expansion of the Universe in the presence of the DM pressure.
The results verify that at lower values of the scale factor, the existence of the
DM pressure leads to the larger DM density.
The Hubble parameter also has higher values when the DM pressure is considered. Our calculations confirm that
the DM pressure results in the increase of the growth rate of
the Hubble parameter with the redshift.
The growth of the scale factor versus the cosmic time is more significant when the DM pressure is present.
In addition, we have shown that the luminosity distance and distance moduli are not considerably influenced by the DM pressure.
Besides, our results indicate that the DM pressure affects the deceleration parameter.

\section*{Acknowledgements}
The author wishes to thank the Shiraz University Research
Council.

\end{document}